# High Temperature Superconductivity in Cuprates: a model

P. R. Silva – Departamento de Física – ICEx – Universidade Federal de Minas Gerais
C. P. – 702 – 30123-970 – Belo Horizonte – MG – Brazil

ABSTRACT

A model is proposed such that quasi-particles (electrons or holes) residing in the $CuO_2$ planes of cuprates may interact leading to metallic or superconducting behaviors. The metallic phase is obtained when the quasi-particles are treating as having classical kinetic energies and the superconducting phase occurs when the quasi-particles are taken as extremely relativistic objects. The interaction between both kinds of particles is provided by a force dependent-on-velocity. In the case of the superconducting behavior, the motion of apical oxygen ions provides the glue to establish the "Cooper pair". The model furnishes explicit relations for the Fermi velocity, the perpendicular and the in-plane coherence lengths, the zero-temperature energy gap, the critical current density, the critical parallel and perpendicular magnetic fields. All these mentioned quantities are expressed in terms of fundamental physical constants as: charge and mass of the electron, light velocity in vacuum, Planck constant, electric permittivity of the vacuum. Numerical evaluation of these quantities show that their values are close those found for the superconducting YBaCuO, leading to think the model as being a possible scenario to explain superconductivity in cuprates.

## 1 – INTRODUCTION

Since the discovery of the high temperature superconductivity in copper oxides (cuprates) by Bednorz and Müller [1], a great amount of theoretical work has been dedicated to understand the mechanism behind this phenomenon. One of the first trying to elucidate this puzzle was proposed by Anderson through the resonant-valence-bond model [2]. Another model [3], due to Emery, assumes that the charge carriers are holes in the O(2p) states and the pairing is mediated by strong coupling to local spin configurations in Cu sites. Emery [3] used an extended Hubbard model in order to describe the main features of this mechanism. Meanwhile Plakida et al [4] explained the high-temperature transition in perovskite-type oxides within the framework of the non-harmonic model for superconductors with structurally unstable lattices. In the model of Plakida and collaborators [4] the highly non-harmonic motion is written in terms of a pseudo-spin representation through a Transverse Ising Model and the interaction of the electrons with the non-harmonic ions vibrations is also described in terms of this pseudo-spin representation.

Two opposite views of the superconductivity in cuprates have been disputed by Anderson and Schrieffer. Anderson [2] attributes the novel phenomenology present on cuprates materials to a second kind of metallic state, namely, the Luttinger liquid. Schrieffer [5] has pursued the interplay between anti-ferromagnetism and superconductivity, extending the BCS pairing theory beyond the Fermi-liquid regime in terms of spin polarons or "bags".

According to Cox and Maple [6] superconductivity in heavy-fermion materials and high-Tc cuprates may involve electron pairing with unconventional symmetries and mechanisms.

As was pointed out by Mourachkine [7]: in1990, Davydov [8,9] presented a theory of high-Tc superconductivity, based on the concept of a moderately strong electron-phonon coupling not treatable by perturbation theory. Also according Mourachkine [7], the theory utilizes the concept of bi-solitons or electron (or holes) pairs coupled in a singlet state due to the local deformation of the –O-Cu-O-Cu- chain in $CuO_2$ planes.

On the other hand, it was shown by Girotti et al [10] that electron-electron bound states are possible in (2+1)-dimensional quantum electrodynamics ($QED_3$). The results obtained in [10] for the relation between the size of the electron pair and the energy of its bound state were interpreted as a possibility of $QED_3$ being a description of high-Tc superconductivity.

A reporter on electron-electron interactions leading to superconductivity in a $YBa_2Cu_3O_7$ (YBaCuO) lattice was published by Harrison [11]. As pointed out by him, the interaction between two electrons near a conducting and polarizable plane is found to be attractive at large separations if the polarizability is large enough. Allowing the phase of the order parameter to vary along the Fermi lines separates the paired electrons sufficiently to sample only the attractive electron-electron interaction.

We would like to cite two other papers in field theory dealing with the high-Tc superconductivity. Belich et al [12] discusses the issue of low-energy electron-electron bound states in Maxwell-Chern-Simons model coupled to $QED_3$ with spontaneous breaking of a local U(1) symmetry. Christiansen et al [13] consider a parity-preserving $QED_3$ model with a spontaneous breaking of a gauged symmetry as a framework for the evaluation of electron-electron interaction potential underlying high-Tc superconductivity.

In1996, we proposed an effective potential as a means to study the mechanism of pairing of high-Tc superconductivity in cuprates [14]. To pursue further on this subject, is the propose of the present work.

In general grounds we think that the basic scenario to develop both conductivity (in the diffusive regime case) as well superconductivity, requires the presence of free electrons (or holes) merged in a fluid of high viscosity. Indeed this could be a macroscopic interpretation for the almost instantaneous establishment of a steady current in metallic conductors as a response to the application of a constant external electric field (d. c. potential). As is well known, superconductivity also requires the interaction of the free electrons (holes) with other degrees of freedom from the lattice. As was pointed out by Maple [6] since electrons repel each other on free space, the pairing (glue) must arise from the solid state. We will describe this interaction through a force linearly dependent on the velocity of the electrons. But as we will see, this force can be traduced in a quantum mechanical description through an interaction contribution to the effective potential.

However we argue that this contribution to the effective potential will be present both in the metallic conductivity case as in the description of the high-Tc superconductivity, the two regimes being distinguished through the free particle contribution to the potential.

## 2 . THE INTERACTION POTENTIAL AND THE REGIME OF METALLIC CONDUCTIVITY

Let us consider that the effective potential describing the metallic (diffusive) conductivity is composed by the sum of two terms. The first of them, the $V_1$ term accounts for the non-interacting part, while the second one, the $V_2$ term, is related to the velocity-dependent force. We write

$$dV_1/dt = -(1/\tau)\,pv. \tag{1}$$

In (1) $p = mv$ is the momentum, with m the mass and v the velocity of the carrier (electron or hole) and $\tau$ is a characteristic time. Upon integrating (1) we get

$$V_1 + (1/\tau)\int pv\,dt = E. \tag{2}$$

The free particle contribution to (2) can be understood through the following reasoning.
a - We consider the non-relativistic kinetic energy of the free electrons (holes) as given by the difference between two virtual energy levels and write

$$h\upsilon_o = p^2/(2m), \tag{3}$$

where h is the Planck constant and $\upsilon_o$ a frequency associated to the virtual transition.
b – Electrons (holes) in motion with the Fermi velocity $v_F$ have the frequency of its emitted virtual photon $\upsilon_o$, Doppler-shifted to $\upsilon_o(1 + v_F/c)$.
c – On the other hand the ions of the lattice could emit virtual photons of frequency $\upsilon_o$ in resonance with the free particles. This frequency also will be perceived by them Doppler-shifted due to their motion.
d – We consider that the information exchanged between the free electrons and the lattice is represented by the beats between the fundamental frequency $\upsilon_o$ and the Doppler-shifted one.
   Taking into account these arguments we can write

$$E_{Met}(p) = (v_F/c)\,[p^2/(2m)] + (1/\tau)\int pv\,dt. \tag{4}$$

It is convenient to translate relation (4), which is expressed in the momentum space in terms of the real-space coordinate R, by using

$$p = h/R, \quad \text{and} \quad v\,dt = dR, \tag{5}$$

we obtain

$$V_{Met}(R) = (v_F/c)[h^2/(2mR^2)] + (h/\tau)\ln(R/R^*), \tag{6}$$

where $R^*$ is a length of reference and $V_{Met}(R)$ is the real-space representation of $E_{Met}(p)$. We minimize (5), putting $dV/dR|_{Ro} = 0$, getting

$$h^2/R_0^2 = (hmc)/(v_F\,\tau). \tag{7}$$

Now we make the identifications

$$h/R_0 = p_F, \quad \text{and} \quad v_F\,\tau = \lambda, \tag{8}$$

where $p_F$ and $\lambda$ are the Fermi's momentum and the mean free path of the electron (hole) respectively. Using (8) into (7) leads to

$$\lambda\,(h/mc) = l_F^2, \tag{9}$$

where h/mc is the Compton wavelength of the electron and $l_F = h/(mv_F)$ is its Fermi wavelength. Relation (9) is equals (except for a numerical factor of order one) to the maximum mean free path of the electron (hole) in the diffusive regime of the electrical conductivity, found in a work of Silva and collaborators [15].

It is possible to evaluate from (5) an effective spring constant k. We have

$$k = d^2V/dR^2|_{R_o} = [(2 p_F^4 v_F)/(h^2 mc)]. \tag{10}$$

In order to get the right side of relation (10) we have also used (7), (8) and (9).

Let us take the Bohr-Sommerfeld method of quantization and write

$$\int_o^1 pdR = h. \tag{11}$$

Evaluating the above integral we have

$$\int_o^1 pdR = <p>(R_1 - R_o) = p_F\, 2A. \tag{12}$$

In (12) we have identified the average momentum $<p>$ with $p_F$, and $R_1 - R_o$ as twice the amplitude of a harmonic oscillator. Comparing (12) and (11) we get

$$A = h/(2p_F). \tag{13}$$

Taking in account (10) and (13) we get $E_{HO}$, the mechanical energy of an equivalent harmonic oscillator. We have

$$E_{HO} = \tfrac{1}{2} k\, A^2 = \tfrac{1}{2} (v_F/c)\, E_F, \tag{14}$$

where $E_F$ is the Fermi energy.

Identifying (14) with $k_B \theta_D$, being $k_B$ the constant of Boltzmann and $\theta_D$ the temperature of Debye, we finally obtain

$$E_F = k_B T_F = 2(c/v_F)\, k_B\, \theta_D. \tag{15}$$

For the gold ($A_u$) a numerical estimation using (15) and $v_F = 1.39 \times 10^6$ m/s (see for instance C. Kittel [16]), gives $\theta_D = 148$ K. This number must be compared with the temperature of Debye of the gold also quoted in reference [16] as 165 K.

3. THE REGIME OF HIGH TEMPERATURE SUPERCONDUCTIVITY

Let us assume that the lattice emits a highly relativistic "virtual" particle of energy pc. This quantum of energy is received Doppler-shifted by the free electrons (holes) in motion with velocity $v_F$ and it beats with its proper frequency of emission in resonance with the lattice. This gives a contribution equal to $v_F p$. Besides this, we consider that this quasi-particle suffers a influence of a force $-p/\tau$ due to the viscosity of the medium. These considerations permit us to write

$$E_{Sup}(p) = E_1(p) + E_2(p) = v_F p + (1/\tau) \int pv\, dt\,. \tag{16}$$

It is also interesting to observe that

$$v_{group|1} \equiv dE_1/dp = v_F. \tag{17}$$

Turning to the real-space representation after use (5), we have

$$V_{Sup}(R) = (v_F h)/R + (h/\tau) \ln(R/R^*). \tag{18}$$

Minimizing (18) with respect to R, we get

$$h/R_0 = h/(\tau v_F) \tag{19}$$

Upon identify $h/R_0$ with $p_F$ we obtain

$$h/\tau = p_F v_F = 2E_F. \tag{20}$$

Also it is possible to evaluate from (18) a new effective spring constant $k_S$. We have

$$d^2V_{Sup}/dR^2|_{Ro} \equiv k_S = p_F^3 v_F/h^2. \tag{21}$$

Working in an analogous way we have done before (see steps (11) to (13) the amplitude of the equivalent harmonic oscillator can be evaluated. Therefore we can write

$$\tfrac{1}{2} k_S A^2 = \tfrac{1}{4} E_F, \tag{22}$$

where we have take in account (13) and (21).
  Thinking in terms of the of the energy equipartition principle it is possible it is possible to maximize the energy of this equivalent harmonic oscillator by imposing the constrain

$$E_F = 4 k_B T_c. \tag{23}$$

We observe that Tc corresponds to the limit of stability of the superconducting phase. It is convenient to write a dispersion relation equivalent to the potential given by (18). We write

$$E_{Sup}(p) = v_F p - v_F p_F \ln(p/p_F). \tag{24}$$

This function exhibits a minimum at $p = p_F$ leading to $E_{Sup}(p_F) = 2E_F$. We observe that

$$v_{group}(p = p_F) = dE_{Sup}/dp|_{pF} = 0. \tag{25}$$

We can interpret $v_g$ as being the velocity of the center of mass of two particles with opposite momentum, $\pm \mathbf{p}_F$. However it seems that the anti-parallelism of the momenta of the two electrons (holes)is not sufficient to warrant the formation of the Cooper pair. An additional interaction between electrons (holes) could be provided by the non-harmonic vibrations of the apical ions [17].
  To pursue further on this subject we got inspiration in the ideas introduced by Chianchi et al [18]. Their phenomenological model for superconductivity in cuprates considers the coupling of holes in $CuO_2$ planes through the mediation of the polarization of the facing AO planes (A= $B_a$, $S_r$) due to the action of the holes themselves. They supposed that the apical site ions perform large amplitude anharmonic oscillations in the **c**-axis direction. In order to construct their model, they followed the Weisskopf intuitive treatment for Cooper pair formation in metals [19].

Let us assume two holes residing in the $CuO_2$ plane with anti-parallel momenta $\pm \mathbf{p}_F$, which approach each other with their linear paths separated by a distance a. At certain instant they collide with a barrier located half-way of the lines representing your paths. This barrier could be due to the motion of the oxygen apical ions, and this inelastic collision conserves the total angular momentum L. As a consequence of the collision the two holes execute circular motion of radius a/2 around this barrier.

Therefore we can write for the holes' motion

$$\omega = 2v_F/a, \quad I = \tfrac{1}{2} ma^2, \quad \text{and} \quad L = I\omega. \tag{26}$$

In (26) $\omega$ is the angular velocity, I the momentum of inertia and L the angular momentum of the pair. The kinetic energy of the pair is given by

$$K_{rot} \equiv L^2/(2I) = mv_F^2. \tag{27}$$

On the other hand, the kinetic energy of a quantum rotor is given by

$$K_Q = L^2/(2I) = \ell(\ell+1)\hbar^2/(2I), \tag{28}$$

where $\ell = 1,2,3,\ldots$, is the angular momentum quantum number.
Making the equality between the ground state of (28) with the right side of (27) leads to

$$v_F = (\sqrt{2}\,\hbar)/(ma). \tag{29}$$

We interpret a, as being the in-plane correlation length which seems to be better realized in cuprates superconductors. We observe that in order a hole pair to occupy the same angular momentum quantum number ($\ell = 1$), they must have anti-parallel spin in order to obey the Pauli's exclusion principle.

We assume that the potential which attracts the pair of holes towards the center of force is of a Coulomb kind, with the holes immersed in a medium of an appropriated dielectric constant. By invoking the "virial theorem", we can write

$$\tfrac{1}{2} mv_F^2 = \tfrac{1}{2} [(4e^2)/(4\pi\varepsilon)](1/a). \tag{30}$$

In (30) $\varepsilon$ is the electric permittivity of the medium and e is hole's electrical charge. The binding energy of the Cooper pair is then given by

$$E_{binding} = \tfrac{1}{2} mv_F^2 - [(4e^2)/(4\pi\varepsilon)](1/a) = -\tfrac{1}{2} mv_F^2. \tag{31}$$

Finally the energy gap is given by

$$E_g = |-\tfrac{1}{2} mv_F^2| = E_F. \tag{32}$$

Indeed the binding energy of the Cooper pair (please see equation (31)) is negative when measured with respect to the minimum of $E_{Sup}(p)$ (please see (24)) considered as the level of zero energy, or in other words, the additional interaction provided by the apical ions lowers the energy of a quantity equals to the energy gap below the minimum of the function $E_{Sup}(p)$.

We can also write by taking in account (23), (29) and (32)

$$E_g = \hbar^2/(ma^2) = 4 k_B T_c \qquad (33)$$

It would been interesting compare eq.(32) with, $E_F = (9/8) E_g$, obtained in [14].

## 4. EVALUATION OF THE CRITICAL CURRENT

We may think that the onset of the critical current $j_c$ will occur when the force provided by the critical internal electric field, namely $e\mathcal{E}_c$, just cancels the force of viscosity of the medium. We write

$$p_F/\tau = e\mathcal{E}_c. \qquad (34)$$

On the other hand we also have the Drude's electrical conductivity $\sigma$ written as

$$\sigma = (e^2 n\lambda)/p_F, \qquad (35)$$

where n is the number of carriers per unit volume and $\lambda$ its mean free path. We have

$$j_c = \sigma \mathcal{E}_c = (en\lambda)/\tau. \qquad (36)$$

Considering eq. (20) and putting

$$n \ell_F^2 \lambda = 1, \qquad (37)$$

where $\ell_F = \hbar/p_F$, we finally obtain

$$j_c = (e v_F p_F^3)/(2\pi\hbar^3). \qquad (38)$$

The critical electrical field can also be determined. By taking in account equations (20) and (34), we get

$$\mathcal{E}_c = (p_F^2 v_F)/(eh). \qquad (39)$$

## 5. THE UPPER CRITICAL FIELD (Hc$_2$)

The present model of superconductivity, which seems to choose the cuprate's lattice as an ideal scenario for its realization, attributes to the holes sited at the $CuO_2$ planes the role of the main players. However when a magnetic field is applied parallel to these planes, we can not rule out circular (helix-form) currents which develop themselves as microscopic solenoids, having the magnetic field as its axis. Next we use this feature to evaluate the critical parallel magnetic field and after we address to the case of the critical field applied in a direction perpendicular to the plane.

### 5a. EVALUATING Hc$_{2\parallel}$

Let us suppose a circular current loop of radius r. The magnetic field $B_{cl}$ at its center is

$$B_{cl} = (\mu_o i)/(2r), \qquad (40)$$

being $\mu_o$ the magnetic permeability of vacuum and i the electrical current. Now we consider an equivalent solenoid of number of turns per unit of length equal to $1/(2r)$. It seems that this solenoid maximizes the uniform magnetic field inside it, and we will take the current as the motion of a Cooper pair. We have

$$B_{sol} = (\mu_o e)/(rT) = (\mu_o e v_F)/(2\pi r^2), \qquad (41)$$

where T is the period of the transverse motion.

We identify $B_{sol}$ with $Hc_{2\parallel}$, when the magnetic energy stored in the solenoid equals to the energy gap. Having in mind this prescription we can write

$$\tfrac{1}{2}(1/\mu_o)(Hc_{2\parallel})^2 \pi r^2 a = E_g = \tfrac{1}{2} m v_F^2. \qquad (42)$$

We can solve (42) for $r^2$ by using (41) and putting $e^2 = 4\pi\varepsilon_o \hbar \alpha c$. We get

$$r^2 = [(\alpha \hbar)/(mc)]\, a. \qquad (43)$$

Therefore we observe that the radius of this micro-solenoid is given by the geometric average between its length (equal to in-plane coherent length of the superconductor) and the classical radius of the electron (hole). Substituting $r^2$ given by (43) into (41) and after use (29) we finally find

$$Hc_{2\parallel} = [(2\sqrt{2})/a^2]\,(\hbar/e). \qquad (44)$$

5b. EVALUATION OF $Hc_{2\perp}$

It is also possible to look at the critical field perpendicular to the $CuO_2$ planes. London's equations which accounts for the exclusion of a magnetic flux from a superconductor can be cast in the form (please see Tilley and Tilley [20])

$$\text{curl } \mathbf{v}_S + (e/m)\mathbf{B} = 0. \qquad (45)$$

We can also write

$$|\text{curl } \mathbf{v}_S| = (e/m)\, Hc_{2\perp}, \qquad (46)$$

where we have identified $Hc_{2\perp}$ with B.

Now, the absolute value of the curl of a vector can be envisaged as the circulation of this vector divided by the area it encloses, in the limit where perimeter and area both shrinks to zero. As we are dealing with a non-continuous medium, we can look at an averaged evaluation of the left side of equation (46). We have

$$|\text{curl } \mathbf{v}_S|_{av} = (v_F/\sqrt{2})[(2\pi l_F)/(\pi l_F^2)], \qquad (47)$$

where $l_F$ is the Fermi's wavelength and the characteristic velocity is took as the Fermi velocity divided by the square root of two. Using the right side of (47) in the left side of (46) and considering that $l_F = h/(mv_F)$, we get

$$m^2 v_F^2 = (eh/\sqrt{2})\, Hc_{2\perp}. \tag{48}$$

Using (29) and solving for the perpendicular critical field we find

$$Hc_{2\perp} = (2\sqrt{2}/a^2)(\hbar/e)(1/2\pi). \tag{49}$$

Finally the comparison between (44) and (49) yields

$$(Hc_{2\parallel})/(Hc_{2\perp}) = 2\pi. \tag{50}$$

## 6. FURTHER DEVELOPMENTS OF THE MODEL AND COMPARISON WITH OTHER RESULTS OF THE LITERATURE

Until now we have essentially considered the motion of the holes belonging to the $CuO_2$ planes (the ab plane). However to better describe the superconductivity in cuprates, the motion of oxygen ions along the **c**-axis perpendicular to these planes must also be considered. Indeed, Cianchi et al attributes in their model [18] an important role to be played by the apical ions. To pursue further on this subject we notice that one of the results of the Landau-Ginzburg (GL) theory has been extended to cover anisotropic materials [21,22], giving

$$Hc_{2\parallel} = \Phi_o/(2\pi\xi_c\xi_{ab}), \tag{51}$$

and

$$Hc_{2\perp} = \Phi_o/(2\pi\xi_{ab}^2). \tag{52}$$

In the above relations $\Phi_o$ is the fluxoid quantum $\xi_{ab} \equiv a$ is in-plane coherence length and $\xi_c$ is the perpendicular (axial) coherence length. Upon dividing (51) by (52) and taking in account (50), we get

$$\xi_{ab} = 2\pi\, \xi_c. \tag{53}$$

We may assume as was considered in a previous work [23] that a pair of holes sitting in the $CuO_2$ plane and circling in phase establishes a ring of charge and that the an oxygen ion could experiment the electric field of this pair of carriers. This leads to an oscillatory motion of the oxygen ion perpendicular to the plane, and in the case of small oscillations, this motion could be described by a harmonic oscillator of frequency $\omega_o$ given by

$$M\omega_o^2 = (4\alpha\hbar c)/\xi_{ab}^3, \tag{54}$$

where M is the mass of the oxygen ion.

Now let us compare the energy of this harmonic oscillator motion performed by the apical oxygen ion with the elementary excitations of the holes related to the energy gap. In order to do this we take into account the anisotropic character of the cuprates and besides this we consider that the Zeeman energy level of a particle is proportional to the magnetic field. With these ideas in mind we write

$$E_{vib}/E_g = (½ M \omega_o^2 z^2)/(h/\tau) = 1/(2\pi) = (H_{c2\perp})/(H_{c2\parallel}), \qquad (55)$$

where $M \omega_o^2$ is the spring constant of the harmonic oscillator and z its amplitude. The factor $2\pi$ comes from (50). Substituting (54) into (50) leads to

$$2\alpha c\tau z^2 = \xi_{ab}^3. \qquad (56)$$

On the other hand, in a similar way we have done before (see eq.((37), we write

$$n\lambda \pi z^2 = [2/(\xi_{ab}^2 \xi_c)] v_F \tau \pi z^2 = 1. \qquad (57)$$

In (56), we took n equal to two holes per coherence volume, and $\pi z^2$ the scattering cross section. Comparing (56) and (55) we finally obtain

$$v_F = (\xi_c/\xi_{ab}) \alpha c /\pi = [\alpha/(2\pi^2)] c. \qquad (58)$$

Relations (58) and (29) permit us to write

$$\xi_{ab} = 2\sqrt{2} \pi^2 [\hbar/(\alpha mc)] = 2\sqrt{2} \pi^2 a_B, \qquad (59)$$

being $a_B$ the Bohr radius.
Comparing (58) and (53) yields

$$\xi_c = \sqrt{2} \pi a_B. \qquad (60)$$

Inserting the value of $\xi_{ab}$ of (59) into relation (33) for the energy gap we obtain

$$E_g = [1/(2\pi^4)](¼\alpha^2 mc^2). \qquad (61)$$

The energy gap relation given by (61) can be thought as the binding energy of a pair of holes of reduced mass equal to half of the free electron mass interacting through a Coulomb potential and immersed in a medium of negative dielectric constant equal to $-\sqrt{2} \pi^2$. The speed of light in this medium being defined as the square root of the product of its magnetic permeability times its electric permittivity will require that it also will have negative magnetic permeability, therefore displaying the effect Meissner.

7. COMPARISON WITH OTHER RESULTS OF THE LITERATURE AND NUMERICAL ESTIMATES

Numerical evaluation of (60) and (59) give respectively 2.35 Å and 14.8 Å for the perpendicular and in plane coherence length. According to Batlogg [24], for the $YBa_2Cu_3O_7$ in the low temperature limit, one find $\xi_{ab} = 14 \pm 2$ Å and $\xi_c = 1.5 – 3$ Å. Results quoted by Burns [22] are respectively 2 – 5 Å and 10 – 20 Å, for the perpendicular and in-plane coherence lengths respectively. The Fermi velocity evaluated from (29) is 1.11 x $10^5$ m/s and could be compared with 1.1 x $10^5$ m/s, as quoted by Batlogg [24] and 1.3 x $10^5$ m/s as estimated by Helmann [25], based in Harrison's work [11]. In a previous paper [14] we found a value of 1.2 x $10^5$ m/s for this quantity. The energy gap given by (60) is estimated as 34.9meV and can be compared with 33.8 meV found in [14].

The critical current density also can be written [after considering (38), (58) and (61)]

$$j_c = [e/(32\pi^9)] (\alpha^4 m^3 c^4)/\hbar^3 = [(2em)/(\pi\hbar^3)] E_g^2. \quad (62)$$

Numerical evaluation of (62), gives 2.48 x $10^{12}$ A/$m^2$ for the critical current, which can be compared with the experimental value of 1.2 x $10^{12}$ A/$m^2$, reported by Kunchur et al [26] for cuprates. Another quotation due to Burns [22] gives the critical current density of 5 x $10^{11}$ A/$m^2$ for $Y_{123}$ films at 4K and for magnetic fields intensities going from 0 to 1 T.

Inserting $\xi_{ab} = a$, given by (58) into the relations for the parallel critical field (44) and the perpendicular critical field we get

$$Hc_{2\parallel} = 1/(2\sqrt{2}\pi^4)[(\alpha^2 m^2 c^2)]/(e\hbar), \quad (63)$$

and

$$Hc_{2\perp} = 1/(4\sqrt{2}\pi^5)[(\alpha^2 m^2 c^2)]/(e\hbar). \quad (64)$$

Numerical evaluations of these magnetic fields give $Hc_{2\parallel}$ = 853 T and $Hc_{2\perp}$ = 136 T. According to Burns [22] these fields are estimated to be at 0 K: 670 T and 120 T, respectively.